# Lattice dynamics and thermodynamics of bcc vanadium at high pressures

Xianwei Sha and R. E. Cohen

Carnegie Institution of Washington, 5251 Broad Branch Road, NW, Washington, DC 20015

We investigate the lattice dynamics and thermodynamics of nonmagnetic bcc vanadium as a function of temperature and pressure, using the first principles linear response linear-muffin-tin-orbital method. The calculated phonon density of states (DOS) show strong temperature dependence, different from inelastic neutron scattering measurements where the phonon DOS show little change from room temperature up to 1273 K. We obtain the Helmholtz free energy including both electronic and phonon contributions and calculate various equation of state properties such as the bulk modulus and the thermal expansion coefficient. A detailed comparison has been made with available experimental measurements.

# I. Introduction

It has been long known that the phonon-dispersion relations of bcc vanadium could not be determined by conventional inelastic neutron scattering techniques since its cross section for neutron scattering is almost totally incoherent. Instead, people use the thermal diffuse scattering of x-rays to measure the phonon frequencies along principle symmetry directions.[1] During the past decade, inelastic neutron scattering techniques have been applied to measure the phonon density of states (DOS) of elemental vanadium as well as its temperature and volume dependence.[2, 3] The phonon DOS of bcc vanadium showed little change from room temperature up to 1273 K and a large softening at 1673 K, which might be attributed to the phonon-phonon scattering.[3] We examined the phonon dispersion and phonon density of states of ferromagnetic bcc Fe using the first-principles linear response linear-muffin-tin-orbital (LMTO) method in the generalized-gradient approximation (GGA), and the theoretical results at both ambient and high pressures show excellent agreements with inelastic neutron scattering data.[4] Applying the same theoretical techniques to bcc vanadium might provide important information to understand the interesting temperature dependence of the phonon DOS.

When using the quasi-harmonic first-principles linear response method to examine various thermal equation of state properties for ferromagnetic bcc Fe, we find that the calculated thermal expansion coefficients agree well with experiment at low temperatures, but the discrepancies increase at high temperatures.[4] One possible reason for the large differences might be magnetic fluctuations at high temperatures, which have not been included in our first-principles calculations yet. We would like to apply the same theoretical techniques to examine the thermal properties of nonmagnetic bcc vanadium where such magnetic fluctuations are absent, to see how the theoretical predictions compare to the experiment at high temperatures.

## II. Theoretical methods

The Helmholtz free energy F for many metals has three major contributions

$$F(V,T)=E_{static}(V)+F_{el}(V,T)+F_{ph}(V,T) \quad (1)$$

with V as the volume and T as the temperature. $E_{static}$ is the zero-temperature energy of a static lattice, $F_{el}$ is the thermal free energy arising from electronic excitations, and $F_{ph}$ is the lattice vibrational energy contribution. We obtain both $E_{static}$ and $F_{el}$ from first-principles calculations directly assuming temperature-independent eigenvalues for given lattice and nuclear positions.[5] The linear response method gives important lattice dynamics information such as the whole phonon spectrum and could be used to obtain $F_{ph}$ within the quasiharmonic approximation. The computational approach is based on the density functional theory and density functional perturbation theory, using multi-κ basis sets in the full-potential LMTO method.[6, 7] The induced charge densities, the screened potentials and the envelope functions are represented by spherical harmonics inside the muffin-tin spheres and by plane waves in the remaining interstitial region. We use the Perdew-Burke-Ernzerhof (PBE) GGA for the exchange and correlation energy.[8] The **k**-space integration is performed over the 16×16×16 grid to construct the induced charge density. We employ the perturbative approach to calculate the self-consistent change in the potential, and determine the dynamical matrix as a function of wave vector for a set of irreducible **q** points on a 8×8×8 reciprocal lattice grid. Careful convergence tests have been made against **k** and **q** point grids and many other parameters to ensure all the results are well converged. We calculate the electronic and phonon density of states at 6 different volumes, 60, 70, 80, 90, 94.6 and 100 $bohr^3$/atom, and obtain the Helmholtz free energies and thermal properties at these volumes and temperatures from 0 to 3000 K in 250 K interval.

# III. Results and Discussions

We show the calculated phonon density of states for nonmagnetic bcc vanadium at five volumes V=70, 80, 90, 94.6 and 100 $bohr^3$/atom in Fig. 1. At the ambient experimental equilibrium volume V=94.6 $bohr^3$/atom, the calculated phonon DOS agrees well with the inelastic neutron scattering measurements[2, 3] at the low phonon frequency end, but not the high frequency end. This is significantly different from bcc Fe, where the calculated phonon DOS shows excellent agreements with the neutron scattering data at both ambient and high pressures.[4] If we use the quasiharmonic approximations where the phonon frequencies at a given volume do not change with temperatures, the temperature dependence of the phonon DOS mainly comes from the thermal expansion. Based on the calculated thermal expansion coefficients, the equilibrium volume at 1273 K is close to 100 $bohr^3$/atom. Given the large differences in the calculated phonon DOS at 94.6 and 100 $bohr^3$/atom and assume the experimental observation of the almost no change in the phonon DOS up to 1273 K[3] is true, there must be some other factors to exert equal and opposite shifts in the phonon frequencies as the thermal expansions with rising temperature, such as the anharmonic effects and the phonon-phonon scattering.

The Helmholtz free energies can be evaluated at different volumes and temperatures from the calculated electronic and phonon density of state. The resulting free energies at each given temperature can be fit to the Vinet equation of state:[9-11]

$$F(V,T) = F_0(T) + \frac{9K_0(T)V_0(T)}{\xi^2}\{1 + \{\xi(1-x) - 1\}\exp\{\xi(1-x)\}\}] \tag{2}$$

where $F_0(T)$ and $V_0(T)$ are the zero pressure equilibrium energy and volume, $x = (V/V_0)^{1/3}$, $K_0(T)$ is the bulk modulus, $\xi = \frac{3}{2}(K_0^{'} - 1)$ and $K_0^{'} = [\frac{\partial K(T)}{\partial P}]_0$. The subscript 0 throughout represents the standard state P= 0 GPa. Pressure can be obtained analytically as:

$$P(V,T) = \{\frac{3K_0(T)(1-x)}{x^2}\} \exp\{\xi(1-x)\} \tag{3}$$

In Fig. 2 we show the calculated presume-volume thermal equation of state at temperatures between 0 to 3000 K. The calculated curve at 250K agrees well with a couple of diamond-anvil-cell x-ray diffraction measurements at the room temperature.[12, 13] The calculated bulk modulus shows a rapid drop with the increase of temperature, as shown in Fig. 3. Kenichi determined the ambient bulk modulus by fitting the experimental pressure-volume data with the Birch-Murnaghan equation of state,[13] and obtained a bulk modulus of 188±2 GPa for fitting of data up to P= 68 GPa, and 162±5 GPa for fitting of data up to 154 GPa. The latter agrees well with our calculated value and earlier x-ray diffraction data.[12]

As shown in Fig. 4, the calculated thermal expansion coefficient α of bcc vanadium shows a linear increase with increasing temperature and a rapid drop with increasing pressure, similar to what has been predicted for bcc Fe.[4] At ambient pressure, the calculated α for bcc vanadium agrees well with experiment[14] up to high temperatures, which is different from the case of bcc Fe. Thermal expansion coefficient is a very sensitive parameter, and the discrepancy between the theory and the experiment for Fe might be attributed to several factors: the errors in the first-principles calculations, anharmonic effects, and most likely magnetic fluctuations for bcc Fe. Since the same theoretical techniques can give much better agreements for bcc vanadium, the contributions of magnetic fluctuations at high temperatures might be one of the major reasons to account for the discrepancies of α in bcc Fe.[4]

# IV. Conclusions

In summary, we have performed detailed first principles linear response calculations to study the lattice dynamics and thermal equation of state properties of nonmagnetic bcc vanadium. The calculated phonon densities of states show a strong change with volume, so there might be other factors to cause the equal and opposite shifts in the phonon frequencies with the thermal expansion, seen in the inelastic neutron scattering experiment, which gave temperature independence of the phonon DOS up to 1273 K. We investigate both the pressure-volume equation of state and the bulk modulus at temperatures between 0 and 3000 K, and the calculated results agree well with the room-temperature x-ray diffraction measurements. The thermal expansion coefficient agrees with the experiment up to high temperatures.

**ACKNOWLEDGEMENTS**

Much thanks to S. Y. Savrasov for kind agreement to use his LMTO codes and many helpful discussions. We acknowledge helpful discussions with B. Fultz. This work was supported by DOE ASCI/ASAP subcontract B341492 to Caltech DOE w-7405-ENG-48. Computations were performed on the Opteron Cluster at the Geophysical Laboratory, supported by the Carnegie Institution of Washington and DOE.

**FIGURES**

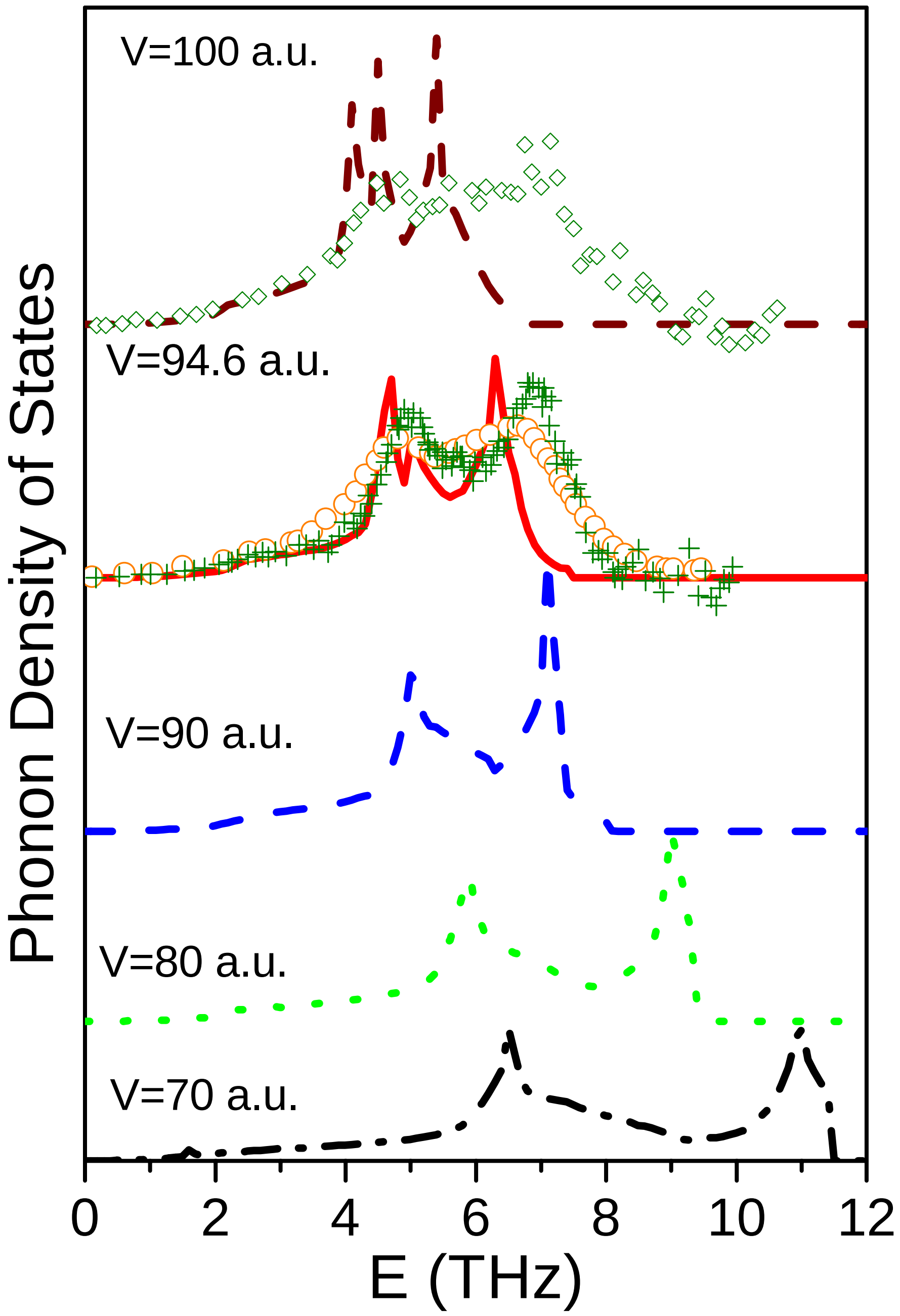

Figure 1. Calculated phonon density of states (lines) for bcc vanadium at different volumes, in comparison to inelastic neutron scattering measurements at room temperature (circles, Ref. 3; cross, Ref. 2 ) and 1273 K(diamonds, Ref. 3). The experimental volumes at ambient temperature and 1273 K are close to 94.6 and 100 $bohr^3$/atom.

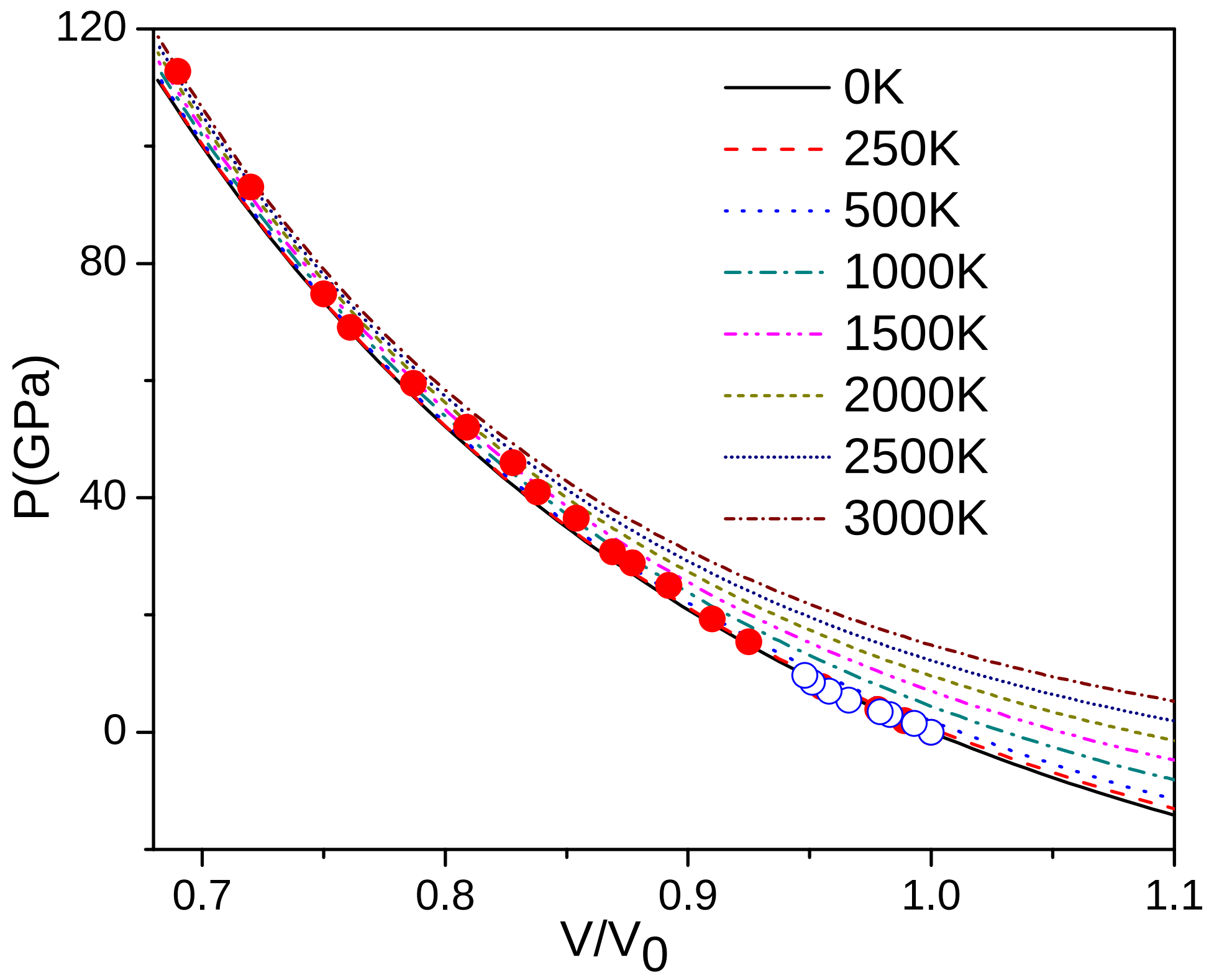

Figure 2. Calculated pressure-volume equation of state for bcc vanadium. The calculated curve at 250 K agrees well with ambient-temperature x-ray diffraction measurements (open circles, Ref. 12; filled circles, Ref. 13)

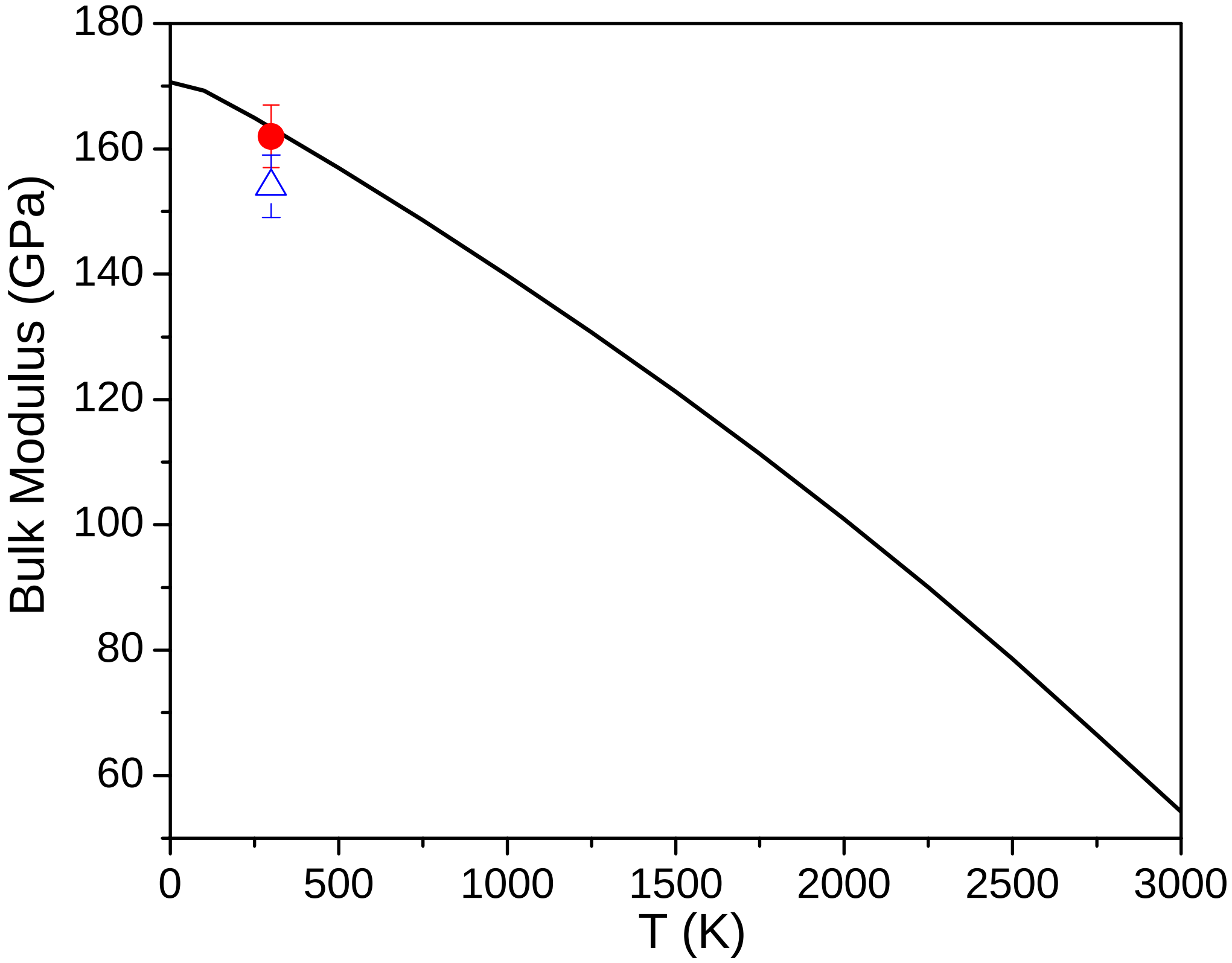

Figure 3. The temperature dependence of the bulk modulus of bcc vanadium. At ambient temperature, the calculated value (the solid line) agrees with the x-ray diffraction measurements (triangle, Ref. 12; circles, Ref. 13).

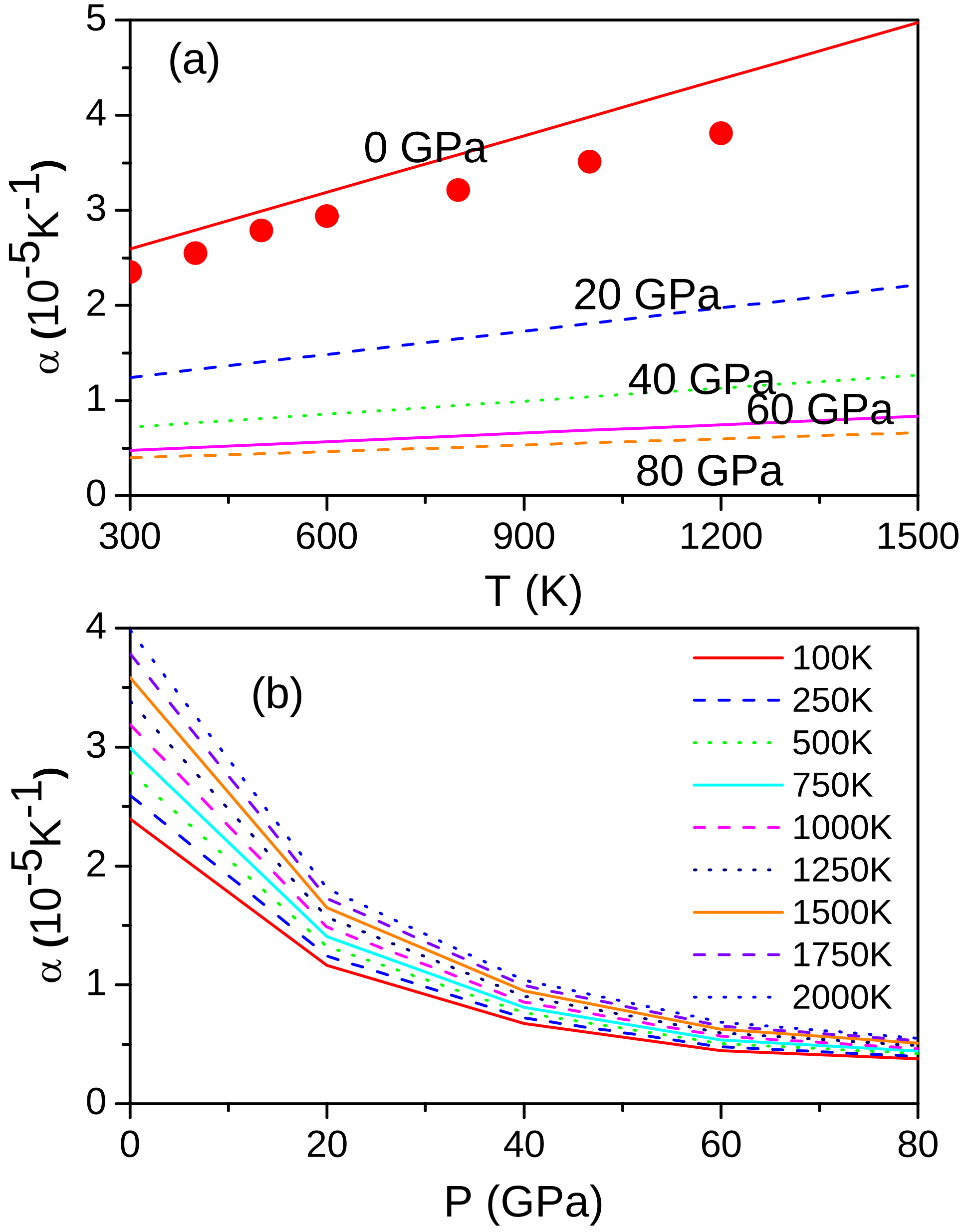

Figure 4. The thermal expansion coefficient of bcc vanadium as a function of temperature (a) and pressure (b). Ambient pressure experimental data (circles) are from Ref. 14.